# Four-body legacy of the Efimov effect


J. von Stecher[1], J. P. D'Incao[1]  & Chris H. Greene[1]

*[1]Department of Physics and JILA, University of Colorado, Boulder, CO 80309-0440*



**Armed with a new solution to the four-body problem, we reanalyze the recent Innsbruck observation of an Efimov resonance in three-body collisions of cesium atoms, and show that it provides evidence of new, universal four-body physics as well.  Our theoretical treatment of four bosonic atoms in the limit of strong two-body attraction suggests a ubiquitous result: namely, that resonantly-bound states of all four atoms should occur whenever the two-body scattering length is a specific fraction (0.43 ± 0.01) of the value where a three-body Efimov resonance occurs. A hitherto unanalyzed feature in the 2006 Innsbruck experiment[1] supports this universal prediction, and it provides the first known evidence to date that a direct four-body recombination process plays a role.  While we confirm the conclusion reached previously by some studies that no true Efimov effect exists for four particles, we demonstrate the existence of a universal class of four-body states that are intimately connected with three-body Efimov states.**


The three-body problem in physics has been a long-standing unsolved challenge for more than a century, dating even farther back than Hilbert's 1900 lecture mentioning it as part of his lecture ``Mathematical Problems''.  Of course Hilbert was alluding to a specific variant of this problem, at that time, in which the three particles are all mutually attracting each other via long-range, inverse-square (gravitational) forces.  In modern times, it is the quantum mechanical version of the three-body problem that is of greatest interest in physics.  While it remains "unsolved", in the sense that no completely general solution has been obtained for all energy ranges, nevertheless tremendous progress has been achieved by both theory and experiment, and it can be argued that most aspects of



the low-energy three-body problem for point particles are becoming well-understood. The 2006 experiment that observed strong evidence for the long-predicted Efimov effect[2,3] justifies this confidence in our theoretical understanding of the three-body problem with short-range forces.

Whereas the three-body problem is beginning to seem ``almost solved'', however, the next step in complexity --- to four interacting particles --- remains in a Neanderthal stage, comparatively speaking. Although a few studies have written down a formal theoretical framework, with which one can imagine attacking the four-body problem, this is largely uncharted territory. Currently, one of the most intriguing question about four-body physics is how it relates to the Efimov effect predicted for *three* strongly interacting bosons. In the early 70's, the nuclear physicist Vitaly Efimov predicted on very general grounds that three neutral bosons whose mutual interaction is characterized by a large value for the two-body *s*-wave scattering length, $a$, can form a large number of weakly bound states whenever $|a| >> r_0$, where $r_0$ is the characteristic range of the interaction.[2,3] Surprisingly, that could happen even when none of the pairs can bind ($a < 0$). For the last 35 years, the desire to observe such Efimov states has been a holy grail of few-body physics and the recent observation of such states in an ultracold atomic gas[1] opens up a new few-body frontier having as a first priority the understanding of the universal properties on the four-boson problem. Here we provide an analysis of this challenging problem which is capable of resolving unsettled questions about the universality in this problem, by demonstrating the existence of a class of universal four-boson states which are intimately related to the Efimov effect. In addition, we show that such universal states can also be seen in ultracold gas experiments in a clear and accessible manner.

One key ingredient of our theoretical model, also present in ultracold gas experiments, is the tunability of the strength of the interactions. For alkali atoms, when



an external *B*-field is placed near a Fano-Feshbach resonance, a small change of *B* can cause *a* to vary from -∞ to ∞, allowing for the exploration of the universal regime. In our model, however, we mimic such variations *a* by explicitly modifying the interatomic interactions[4,5]. In our framework, the solution of the four-body problem culminates with the solution of the ``hyperradial'' Schrödinger equation:

$$-\frac{\hbar^2}{2m}\frac{d^2}{dR^2}F_\nu(R) + \sum_{\nu'} W_{\nu\nu'}(R)F_{\nu'}(R) = EF_\nu(R) \qquad (1)$$

where the hyperradius *R* describes the overall size of the system. Here, *m* is the atomic mass, *E* the total energy, $F_\nu(R)$ the hyperradial wave function with $\nu$ representing the set of quantum numbers needed to label each channel. In the hyperspherical adiabatic representation[4,5] most of the complexity of the problem arises in solving the hyperangular equations to determine the effective potentials $W_{\nu\nu}(R)$ and couplings $W_{\nu\nu'}(R)$, in this case using an innovative numerical technique (von Stecher J. & Greene, C. H. in preparation). The reduction of the problem to the hyperradial Schrödinger equation (1) then leads to a simple and intuitive picture: the effective potentials $W_{\nu\nu}(R)$ support all bound and quasi-bound states of the system, while the off-diagonal nonadiabatic couplings $W_{\nu\nu'}(R)$ drive inelastic transitions among different channels.

We explore the universality of the four-boson and its relation to Efimov physics by solving Eq. (1) for the energy spectrum across the relevant regime of interaction strengths *a*. This spectrum is summarized in Fig. 1(a), where the four-bosons energies (solid black lines) are obtained using two complementary numerical techniques: the adiabatic hyperspherical approximation,[6,7] and the correlated Gaussian basis set expansion[8,9] (see Supplementary Information for details). Note that, in order to visualize the full energy landscape over a broad range of *a*, we have introduced the auxiliary



function $F(x) \equiv \text{sgn}(x)\ln(1 + |x|)$. This figure, although somewhat busy, shows all the important features that relate two-, three-, and four-boson physics. Figure 1(a) includes the possible dimer-dimer and dimer-atom-atom breakup thresholds for $a>0$ (red solid lines) as well as the energies of the Efimov states (green dashed lines) representing the atom-trimer breakup threshold. Figures 1(b) and 1(c) convey the geometrical nature of these states taking into account the extremely ``floppy" nature of the Efimov trimer states in which all possible triangular shapes, and even linear configurations are comparably probable[4,5,10].

Figure 1(a) implies that the four-boson spectrum, throughout the range $r_0/|a| << 1$, is characterized by precisely two tetramer states that are associated with each three-body Efimov state. This quantitatively confirms a conjecture to this effect in Ref.12.   In fact, our extensive numerical tests show that these four-boson energies obey a universal relationship to the corresponding Efimov state energy, which at unitarity ($|a|=\infty$) can be expressed as

$$E_{4b}^{(n,m)} = c_m E_{3b}^{(n)}, \qquad\qquad (2)$$

where $E_{3b}^{(n)}$ is the energy of the $n$th Efimov state, $n$=0,1,2,..., and $E_{4b}^{(n,m)}$, $m$=1 and 2, are the two tetramer energies associated with it (see Supplementary Information for details). The universal relation between three- and four-body energies has been found to be characterized by these two *universal numbers*, equal to: $c_1 \approx 4.58$ and $c_2 \sim 1.01$. This universal relation among three- and four-body energies has further implications. Specifically, the three-body Efimov effect exerts a pervasive influence on the four-boson system, in that the four-boson energies follow precisely the same geometric scaling as the three-boson energies, with successive energies related by the factor $e^{-2\pi/s_0} \approx 1/515$ and successive radii expanding by the factor $e^{\pi/s_0} \approx 22.7$ (where $s_0 \approx 1.00624$).



However, there are limits to the types of states for which this universal picture applies. When the Efimov state is sufficiently compact, as is usually the case for the Efimov ground state or lower trimer states with no Efimov character at all, the details of the interatomic interaction substantially modify the relationship between three- and four-body energies. For instance, for the energies in Fig. 1(a) we obtained : $c_1 \approx 5.88$ and : $c_2 \approx 1.10$. On the other hand, for long-range Efimov states, as is certainly the case of excited Efimov states, we have verified that the relation (2) is truly universal, i.e., it holds independently of the model adopted for the two-body interactions, to within better than 2% accuracy. In fact, we believe that the ability of the current method to calculate many more weakly-bound energy levels than previous techniques has been decisive, and it permits us to resolve a previously-existing disagreement in the literature, between the results of Hammer and Platter *et. al*[11,12] and those of Yamashita *et. al*[13]. The unified picture that has emerged from our calculations now resolves this controversy: it is related to the fundamental question of whether or not an additional ``four-body parameter'', encapsulating nonuniversal aspects from the details of the interactions, is required to specify the nature of the four-boson spectrum and scattering observables, akin to the usual three-body parameter[14].

Specifically, we support the qualitative conclusion of Hammer and Platter[11,12] that no additional four-boson parameter is required, however, *only* for states associated with long-range Efimov trimers, i.e., when the universal regime is quantitatively achieved. This result also explain the observations of Yamashita *et. al*[13] that the energies of more compact four-boson states can vary depending on the details of the interatomic interactions. Our understanding emerges from Fig. 2, showing the four-body effective potentials calculated at unitarity, $|a|=\infty$. We have verified that the minimum of the effective potentials in fact scales with the size of the trimer state as indicated by vertical dashed lines in Fig. 2. Therefore, if the lowest Efimov state has a size that exceeds $r_0$ only marginally, the minimum of the four-body potential is close enough to $r_0$ to allow



the four-body states to access the nonuniversal short-range physics. On the other hand, if the lowest Efimov state is large compared to $r_0$, then the minimum of the four-body potential lies at $R \gg r_0$ and the four-boson states probe almost no nonuniversal effects. These universal properties allow us to point out that the four-boson states are truly universal, obeying the same geometric scaling as the Efimov states; thus, for this extensive class of states, no additional four-body parameter is required.

Parenthetically, we have in addition confirmed the existence of a class of four-body states that are simply a realization of the Efimov effect for three-bodies. As pointed out by Braaten and Hammer[15], these four-atom states occur whenever a Efimov state is created for $a > 0$ (see the regions in $a$ marked with a square). In this case the atom-dimer scattering length, $a_{ad}$, goes though a pole and whenever $|a_{ad}| \gg a$ a series of four-body Efimov states, which can be seen as a dimer-atom-atom state, appear as a result of the emergence of an attractive dipole-like potential ( $\propto -1/R^2$ ) in the dimer-atom-atom channel, for $a \ll R \ll |a_{ad}|$. Our results also confirm the predictions of Amado and Greenwood[16] on the non-existence of a similar Efimov effect for four-bosons. That can be simply seen from Fig. 1(a) where in the threshold for formation of the trimer states for $a < 0$ there will exist only a finite number of four-body states.

Here we also demonstrate that the universal four-body physics we discussed above can readily be observed in ultracold quantum gases. In general, weakly bound states can deeply affect the collisional properties of ultracold gases enhancing the atomic and molecular losses. The relative importance of four-body processes, however, remains largely unexplored and one could argue that such processes should be far less likely than two-body or three-body collisions in a typical low-density gas. On the other hand, near the threshold for formation of any four-boson states, the four-body scattering observables should exhibit a resonant enhancement that can dramatically affect the collisions, even for typical low densities. The results of the Innsbruck experiment[1],



realized at atomic densities of $n(0) \approx 3 \times 10^{12}$ cm$^{-3}$, were interpreted under the premise that the atom loss is entirely due to three-body recombination, a process where three atoms collide to form a bound dimer molecule and a free atom, which releases enough kinetic energy to cause the collision products to escape from the trap, $B+B+B \rightarrow B_2+B+E_{\text{rel}}$. Sure enough, the experimental data shows a clear resonant peak on the three-body recombination rate $K_3$, more specifically at a three-body recombination length of $\rho_3 = \left[2mK_3/(\sqrt{3}\hbar)\right]^{1/4}$, at $a$=-850 a.u. in excellent agreement with theoretical expectations for the manifestation of Efimov physics through three-body recombination[14,17-22].

Although the experimental data for the Innsbruck experiment are reasonably well-understood[14,17-22], the task of distinguish three- and four-body losses is extremely difficult, and four-body processes could still be embedded in the observed decay rates. Accordingly, we have reanalyzed the experimental data from the Innsbruck experiment, looking for possible signatures of new four-boson states. The key observation from our results is that for $a$<0, when an Efimov state is created, say at $a = a_{3b}^*$, it is accompanied by the creation of two four-boson states at slightly less negative values of $a$, and those states are expected to enhance four-body processes for the region of $a$ the experimental data was taken. Moreover, our calculations indicate that once the scattering length $a_{3b}^*$ is known, one can determine the scattering lengths in which such four-body states appear. This universal relation can be simply determined by the energy spectrum [see Fig. 1(a)] and it is expressed as

$$a_{4b1}^* \approx 0.43 a_{3b}^*, \text{ and } a_{4b2}^* \approx 0.90 a_{3b}^*. \quad (3)$$

The main four-body process in which such four-boson states should be manifested is four-body recombination, where the four initially free atoms collide to recombine into



the dimer-dimer channel and/or into the atom-trimer channel. Figures 3(a) and 3(b), respectively, depict this process through the effective potentials at scattering lengths very close to the threshold for formation of an Efimov state (green dashed line) and just after its creation. When a four-boson state resides in energy close to the collision threshold, one expects a resonant enhancement to the four-body recombination rate, $K_4$. A straightforward Wigner threshold-law analysis demonstrates that $K_4$ approaches a constant as the collision energy is tuned to zero, and thus four-body recombination can indeed potentially compete with three-body recombination in causing atomic losses (N. P. Mehta, S. T. Rittenhouse, J. P. D'Incao, and C. H. Greene. A general theoretical description of N-body recombination, unpublished).

In order to assess the importance of $K_4$ and quantify our predictions, we have calculated $K_4$ by numerically solving Eq. (1) using a formula for $K_4$ derived elsewhere (N. P. Mehta, S. T. Rittenhouse, J. P. D'Incao, and C. H. Greene. A general theoretical description of N-body recombination, unpublished). The main difficulty in comparing our results to the experimental data is that the existing experiments are probably unable to distinguish three- and four-body losses. We therefore introduce an *effective* three-body recombination rate, in which both three- and four-body physics are included. We define it as,

$$K_3^{eff}(a,t) = K_3(a) + n(t)K_4(a)/3,\qquad(4)$$

where $n(t)$ is the atomic density at any chosen time *t,* calculated by solving the rate equations for its time evolution. We have in fact verified that for times $t<\sim t_0=[n(0)^2(K_3+n(0)K_4)]^{-1}$ ($\approx$50 ms for our case) the time dependence of $n(t)$ can be described as result of the effective three-body rate in Eq. (4), by setting $n(t)=n(0)$. For



longer times, $n(t)$ can result from three- or four-body processes depending on whether or not $K_3 >> n(0)K_4$. Our results for $\rho_3^{\text{eff}} = \left[2mK_3^{\text{eff}} /(\sqrt{3}\hbar)\right]^{1/4}$ are shown in Fig. 3(c) for

$t$=20 ms. For $K_3$ we use the thermally averaged results of Ref. [14] calculated for

temperatures of 10, 200 and 250nK, and adjust it to fit the Efimov resonance at

$a = a_{3b}^{*}$ = -850 a.u. and the experimental data for $|a| > |a_{3b}^{*}|$. Our results show that for

this range of $a$ three-body recombination is indeed the dominant loss process. For

$|a| < |a_{3b}^{*}|$, however, we find much better agreement by assuming that four-body

recombination is the dominant loss process --- the dashed curve in Fig. 3(c) is the

contribution from $K_3$ at 10nK. For this range of $a$, as indicated in Fig. 3(c) by the

vertical dashed lines at $a = a_{4b1}^{*}$ and $a = a_{4b2}^{*}$, $K_4$ is resonantly enhanced when the two

four-boson states are created [see region in Fig. 1(a) marked by a circle]. This

agreement strongly suggests that the Innsbruck experiment also offers the first

experimental evidence for the universal four-boson states we discussed here, although

the agreement with the second resonance predicted for $a = a_{4b2}^{*}$ and 10nK requires

some imagination --- and for temperatures 200 and 250nK this resonance feature is

washed out due to thermal effects. Nevertheless, the verification of the universal

constraint between three- and four-body physics as given in Eq. (2) strengthens even

more our conclusion that the main resonant feature at -850a.u. is caused by Efimov

resonance.

Note also that for $|a| > |a_{3b}^{*}|$, four-body recombination to Efimov states,

$B+B+B+B \rightarrow B_3+B$ [see Fig. 3(b)], is likely to be the dominant decay pathway.

Although three-body recombination tends to dominate the atom loss, the formation of

Efimov states through four-body recombination is not negligible. In Fig. 3(d) we show

the atomic density ($n_B$) and the density of trimers $n_{B_3}$ at 10nK for $a$ up to -10000a.u..

Near the threshold for formation of the Efimov state, the energy released through four-



body recombination is small (approximately the trimer binding energy) and the Efimov state can remain trapped. In this case, our results indicate that about 10% of the atoms will form trimers. For larger $|a|$, however, the trimer formation is strongly enhanced by a four-body resonance associated to the lowest four-boson state attached to a second Efimov state [see Fig. 1(a)]. In this case, we find that about 50% of the atoms will form trimers. The drawback, however, is that the energy released through four-body recombination can make both atom and the Efimov trimer to escape from the trap. Nevertheless, in an experiment where only atoms are visible, one could set the magnetic field to a value such that $|a| > |a^*_{3b}|$, measure the number of atoms and tuning field back to a value such that $|a| < |a^*_{3b}|$ where no trimer exist. The reappearance of atoms after this process would be a convincing signature of the first experimental realization of an ultracold gas of Efimov trimers.

**Summary**

While there is no ``true Efimov effect'' for four-bosons, this does not mean that Efimov physics is irrelevant for the four-boson problem. Remarkably, many of the details of short-range atom-atom interactions are inconsequential, and the quantum states for 4 neutral bosonic particles obey simple, universal scaling relations as expressed in terms of their sizes, energies, and the scattering lengths at which they appear. Moreover, a hitherto uninterpreted feature in the Innsbruck experiment[1] can be assigned through our unified theoretical framework as a four-body bound state that is reflected in a four-body recombination loss process. This Article also presents, apparently, the first prediction of the absolute rate for four-body recombination in an ultracold quantum gas.

**Acknowledgments** This work was supported in part by the National Science Foundation. We are indebted to N. Mehta and S. Rittenhouse for extensive discussions and for access to their unpublished




derivations prior to publication. We also thank the F. Ferlaino, S. Knoop, H.-C. Nägerl, and R. Grimm from the Innsbruck group for the discussions on their experimental data.



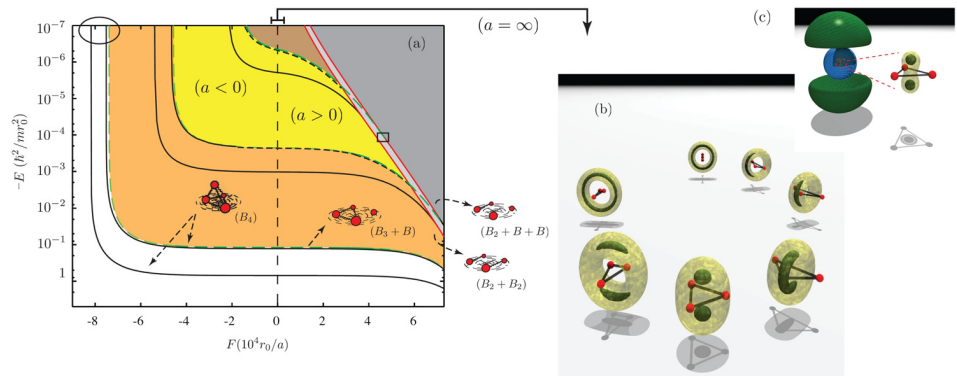

**Figure 1| Spectrum of energies and geometric structure of four-boson states and their connection to the Efimov physics**. In (a) we show the energy spectrum of four-bosons as a function of *a* using the auxiliary function $F(x) \equiv \mathrm{sgn}(x)\ln(1+|x|)$ designed to aid the visualization of the full energy landscape. Black solid lines correspond to four-body states, dashed green lines represent atom-trimer dissociation thresholds, while red lines correspond to dimer-atom-atom (upper) and dimer-dimer (lower) thresholds. (b) and (c): geometrical nature of ground and excited four-body states, respectively, for different trimer configurations. For the four-boson ground state, the isosurfaces displayed are those where the probability of finding the fourth atom is 0.9 and 0.99 of the maximum probability for that particular trimer geometry. The probability isosurface for the first excited four-boson state in the portion of space where the Efimov trimer resides at its most probable equilateral triangular geometry implies that the fourth atom is very weakly bound, and its size exceeds that of the ground state considerably.



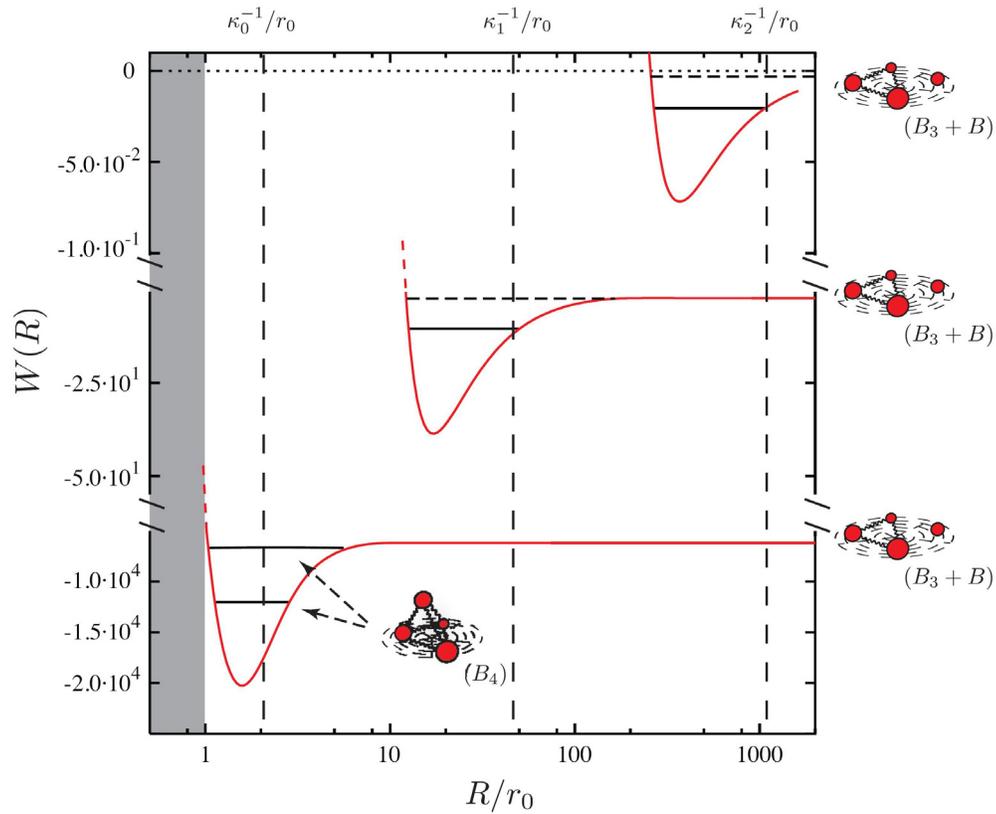

**Figure 2| Effective four-boson potentials for |*a*|=∞ converging at large *R* to the atom-trimer thresholds.** Solid and dashed black lines represent the four-boson states shown in Fig. 1(a). The position in *R* of the minimum of these potentials scales with the size of the Efimov state, indicated in the figure by $\kappa_i^{-1}$ (see Supplementary Information for definition). Therefore, when the trimer state is large the four-boson states associated with it will lie at large *R*, preventing access to the nonuniversal region *R* <~ *r₀*, whereby the four-bosons state are universal.



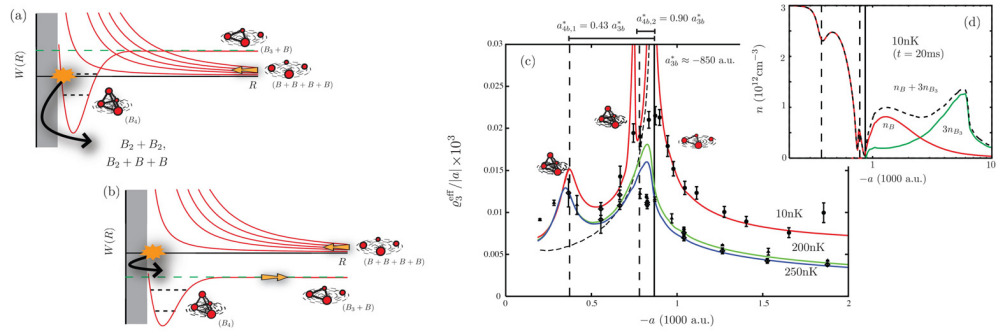

**Figure 3| Evidence of signatures of four-boson states through four-body recombination.** In (a) and (b) we schematically represent the pathways for four-body recombination when the Efimov state (green dashed line) is above and below the four-body break-up threshold, respectively. In (a) the only possible decay channels are associated with deeply bound (two- and three-body) states, while in (b) the main decay channel is to the available Efimov state plus a free atom. When a four-boson state crosses the four-body dissociation threshold, it resonantly affects the four-body recombination rate, enhancing the rate of atom losses in an ultracold gas. (c) Comparison of $\rho_3^{\text{eff}}$ [Eq. (4)] with the experimental data from the Innsbruck group[1]. The vertical lines identify the critical scattering lengths where an Efimov state and its associated four-body states are created, respectively, $a_{3b}^* \approx$ -850 a.u., $a_{4b1}^* \approx 0.43 a_{3b}^* \approx$ -365.5 a.u., and $a_{4b1}^* \approx 0.43 a_{3b}^* \approx$ -765 a.u.. In (d) we show the atom and trimer densities at 10nK, demonstrating that four-body recombination can be used to form Efimov trimers.